\documentclass{Interspeech2024}

\interspeechcameraready 

\usepackage{cite}

\title{The Interspeech 2024 Challenge on Speech Processing Using Discrete Units}

\name[affiliation={1}]{Xuankai}{Chang}
\name[affiliation={1}]{Jiatong}{Shi}
\name[affiliation={1}]{Jinchuan}{Tian}
\name[affiliation={4}]{Yuning}{Wu}
\name[affiliation={4}]{Yuxun}{Tang}
\name[affiliation={4}]{Yihan}{Wu}
\name[affiliation={1}]{Shinji}{Watanabe}
\name[affiliation={2}]{Yossi}{Adi}
\name[affiliation={3}]{Xie}{Chen}
\name[affiliation={4}]{Qin}{Jin}

\address{
  $^1$Carnegie Mellon University, USA \, $^2$The Hebrew University of Jerusalem, Israel \\
  $^3$Shanghai Jiao Tong University, China \, $^4$Renmin University of China, China}
\email{xuankaic@andrew.cmu.edu,jiatongs@andrew.cmu.edu}


\begin{document}

\maketitle

\begin{abstract}
    Representing speech and audio signals in discrete units has become 
    a compelling alternative to traditional high-dimensional feature vectors.  
    Numerous studies have highlighted the efficacy of discrete units in various applications such as speech compression and restoration, speech recognition, and speech generation. To foster exploration in this domain, we introduce the Interspeech 2024 Challenge, which focuses on new speech processing benchmarks using discrete units. It encompasses three pivotal tasks, namely multilingual automatic speech recognition, text-to-speech, and singing voice synthesis, and aims to assess the potential applicability of discrete units in these tasks. This paper outlines the challenge designs and baseline descriptions.
    We also collate baseline and selected submission systems, along with preliminary findings, offering valuable contributions to future research in this evolving field.
\end{abstract}

\section{Introduction}
\label{sec:intro}

In the realm of automatic speech recognition (ASR), considerable advancements have unfolded in the past few decades, propelled by the emergence of deep neural networks~\cite{hinton2012deep,qian2016very}. Recently, the predominant approach has shifted towards end-to-end (E2E) ASR models~\cite{graves2006connectionist,graves2012sequence,chorowski2015attention}, gaining popularity and witnessing performance enhancements through a spectrum of robust architectures~\cite{dong2018speech,gulati2020conformer,guo2021recent,kim2023branchformer}. Noteworthy strides have also been made in training methodologies, with self-supervised learning (SSL) models~\cite{baevski2020wav2vec,hsu2021hubert,chen2022wavlm} and large-scale supervised training, such as Whisper~\cite{radford2023robust}, demonstrating improved performance and generalization. Traditionally, high-dimensional features are derived from raw waveforms in most endeavors. Spectral speech features, like Mel Frequency Cepstral Coefficients (MFCC) or log Mel filter banks (FBANK), conventionally stem from fixed-length temporal windows. Recently, learnt features based on deep neural networks through data-driven methods have become mainstream~\cite{sainath2015learning,baevski2020wav2vec,hsu2021hubert}. Despite these innovations, the data storage and transmission efficiency remain comparable between raw waveforms and speech features in many cases~\cite{chang2023exploration}. The challenge persists in enhancing computational efficiency without compromising performance integrity.

Recently, there has been a surge in the adoption of discrete speech representations, with notable developments in the Generative Spoken Language Model (GSLM) for textless Natural Language Processing (NLP) exemplified by \cite{lakhotia2021generative, kharitonov2021text}. GSLM leverages techniques akin to those used in language modeling to address speech processing tasks through discrete speech representations. The representation of speech as discrete tokens presents a unique advantage, allowing for the unified modeling of both speech and text data within a streamlined framework. Several previous studies have highlighted the efficacy of jointly modeling speech-text data, showcasing improved performance in tasks related to speech and text generation \cite{rubenstein2023audiopalm,maiti2023voxtlm, hassid2023textually}. Moreover, employing manipulation methods on discrete tokens enables the reduction of sequence length, resulting in more efficient computation \cite{lakhotia2021generative,chang2023exploration}.

To encourage further exploration in this field, we propose the challenge of ``Speech Processing Using Discrete Speech Units''. The significance of this topic lies in its transformative potential across various applications within the community of speech and natural language processing~\cite{kreuk2021textless, nguyen2023generative, maimon2022speaking, hayashi2020discretalk, kim2023many, kreuk2022audiogen, copet2023simple, hsu2023revise}. The primary goal of this challenge is to advance innovation and investigation in the domain of discrete speech units, a field that has recently showcased remarkable potential but still lacks unified evaluation platforms to benchmark these methods.
To fulfill this objective, we outline three core tasks:
\begin{enumerate}
    \item The ASR task focuses on the multilingual aspect by incorporating data from the ML-SUPERB challenge \cite{shi2023ml}.
    \item The TTS task is divided into two tracks: a \textit{single-speaker TTS track}, which focuses on synthesizing speech from text using a single voice, and a \textit{vocoder track}, which concentrates on the resynthesis of expressive, multi-speaker speech.
    \item The SVS task focuses on synthesizing single-singer singing from musical score information.
\end{enumerate}
We chose these tasks due to their broad applicability and established benchmarks, which ensure clear evaluation metrics and significant real-world impact. These tasks cover the complete speech processing pipeline, encouraging holistic innovation in discrete unit processing. Additionally, they reflect current research trends and present diverse challenges that thoroughly test the capabilities of discrete unit representations, driving meaningful advancements in the field.
This paper details the challenge designs, baselines, and evaluation metrics with ranking, which consist of ASR/TTS/SVS performance measures and compression rates. 
In addition, we provide preliminary analyses, including both baselines and selected results submitted at this juncture, to help us find new research directions. 

\section{Challenge Details}
\label{sec:challenge_details}

\subsection{Formulation of discretization and bitrate}

We denote an input waveform with $T$ sampled data points with a sampling rate $S$ as $\mathbf{x} \in \mathbb{R}^T$. 
This challenge defines discretization $f(\cdot)$ as a function to project $\mathbf{x}$ into a set of discrete sequence streams $\mathbf{U} = \{ U^1, \ldots, U^M \}$, where we allow $M$ streams and $U^m$ denotes the $m$\textsuperscript{th} stream of discrete tokens. 
The $U^m$ is defined as $U^m = (u_i^m \in \mathcal{V}_m | 1 \leq i \leq N_m)$, where $N_m$ and $\mathcal{V}_m$ are the sequence length and the vocabulary/codebook of the $m$\textsuperscript{th} stream, respectively.

Based on this formulation, we define the bitrate $B$ (bit/second) of the discrete representation $\mathbf{U}$ given the original waveform samples length $T$ and its sampling rate $S$:
\begin{equation}
\label{eq: bitrate}
    B = \sum_{m=1}^M \left(\frac{N_m \cdot \log_2(|\mathcal{V}_m)|}{T/S}\right),
\end{equation}
which corresponds to the sum of the bitrates from all levels.
Bitrate is an important metric in our challenge to measure the efficiency of discrete representation.

\subsection{ASR task}
\label{ssec:asr_track}
To assess the fidelity of semantic information, we incorporate the ASR track in the challenge.

\noindent \textbf{Task Definition and Baseline}: The target of ASR is to transcribe speech signals into text. Traditionally, feature extraction is applied to an audio segment of length $W$ ($W \leq T$), represented as $\mathbf{x}_i = \mathbf{x}\left[t_i:t_i+W\right]$, undergoes conversion to a $D$-dimensional vector of real or complex values, denoted as $\mathbf{X}_i \in \mathbb{C}^D$. Here, $\mathbf{X}_i$ signifies the feature of that segment, commonly referred to as a frame. In the context of ASR tasks utilizing discrete units, the feature of a frame, $\mathbf{X}_i$, is represented as $U_i = \{ u^1_i, \ldots, u^M_i \}$. Notably, in certain instances, as seen in \cite{chang2023exploration}, $M=1$ is employed. In such cases, the sizes of $\mathbf{X}_i$ and $u_i$ are $32 \times D$ and $\log_2(|\mathcal{V}|)$ bits, respectively, under the assumption that $\mathbf{X}_i$ is stored in 32-bit float value and $|\mathcal{V}|$ denotes the size of the codebook.

In the study conducted by Chang \textit{et al.}~\cite{chang2023exploration}, it was established that discrete units-based ASR systems exhibit proficient performance on the majority of mono-lingual datasets. However, challenges arise in the context of multi-lingual scenarios, as evidenced by the ML-SUPERB dataset~\cite{shi2023ml}. Consequently, this challenge deliberately emphasizes and promotes the multi-lingual dimension of discrete units-based ASR.

\noindent \textbf{Data}: To stress the multi-lingual aspect mentioned above, in addition to the widely-used LibriSpeech~\cite{panayotov2015librispeech} 100-hour subset (LibriSpeech-$100$), we also adopt the ML-SUPERB 1-hour public benchmark~\cite{shi2023ml} in the ASR task. LibriSpeech-$100$ comprises a clean, read English corpus, effectively addressed by the discrete units-based ASR~\cite{chang2023exploring}. In contrast, ML-SUPERB presents a more formidable challenge, given the complexities of language families with $143$ languages and the limited volume available for each language. 
Notably, the $1$-hour track from ML-SUPERB encompasses approximately $220$ hours of speech.
The training sets of both corpora are combined to train the ASR model, with the inclusion of LibriSpeech-$100$ aimed at easing training complexities and showcasing performance on a data-rich resource. As for the evaluation, we employ all test sets from LibriSpeech (dev-clean, dev-other, test-clean, and test-other) and ML-SUPERB (test\_$1$h). The data preparation scripts are included in the baseline by following conventional methods of LibriSpeech-$100$ and ML-SUPERB. It is important to highlight that there are no constraints on the data used for obtaining discrete tokens in this challenge, including pre-training and $k$-means training.

\noindent \textbf{Evaluation Metrics}: Two evaluation metrics are employed: Character Error Rate (CER) and bitrate.
\begin{itemize}
    \item CER: The test sets are categorized into two groups, encompassing English and multi-lingual content. Consequently, two CERs are computed: $\text{CER}_\text{EN}$ and $\text{CER}_\text{ML}$. $\text{CER}_\text{EN}$ is calculated across all utterances in the LibriSpeech test sets, while $\text{CER}_\text{ML}$ is computed on the ML-SUPERB test set. The adoption of CER for LibriSpeech ensures consistency with ML-SUPERB.
    \item Bitrate: The calculation follows Eq.~\eqref{eq: bitrate}. We compute the bitrate on the whole test sets, i.e., all librispeech evaluation sets and ML-SUPERB test sets. 
\end{itemize}

\noindent \textbf{Ranking}:
The overall ranking is based on the average of all three ranking positions: \begin{itemize}
  \item $R_1$: micro average $\text{CER}_\text{EN}$ on all LibriSpeech test sets;
  \item $R_2$: $\text{CER}_\text{ML}$ on the ML-SUPERB test set;
  \item $R_3$: the bitrate of the overall test sets.
\end{itemize}
The overall ranking position is $\hat{R} = \frac{(R_1 + R_2 + R_3)}{3}$. 
In cases where multiple systems share the same ranking, the tiebreaker is determined by the order $R_2 > R_1 > R_3$.

\subsection{TTS (Vocoder) task}
\label{ssec:tts_vocoder_track}

In the TTS (vocoder) track of the challenge, the focus is on the conversion of discrete speech units into waveforms, assessing the acoustic information within these units.

\noindent \textbf{Task Definition}: The core objective of vocoder modeling (speech resynthesis) is to develop a reverse function $f^{-1}(\cdot)$ capable of transforming discrete speech units $\mathbf{U}$ into an audible waveform $\hat{\mathbf{x}}$. No restrictions are placed on the type or size of the model used for the vocoder.

\noindent \textbf{Data}: The dataset for this task is sourced from the Expresso benchmark~\cite{nguyen2023expresso}, focusing solely on single-speaker scenarios to avoid complications with multi-speaker conversions and long-form speech. The data is partitioned into training ($9.7$ hours), development ($0.6$ hour), and test ($0.6$ hour) sets, and while discrete unit learning can utilize external data, vocoder training is restricted to the provided training dataset.

\noindent \textbf{Evaluation metrics}: Four metrics are employed for evaluation: Mel cepstral distortion (MCD), F0 root mean square error (F0 RMSE), UTMOS~\cite{saeki2022utmos}, and bitrate. UTMOS is calculated using the winner model from the VoiceMOS 2022 challenge~\cite{huang22f_interspeech}, and the bitrate calculation is standardized as Eq.~\eqref{eq: bitrate}. The evaluation process is facilitated by ESPnet-TTS~\cite{hayashi2020espnet, hayashi2021espnet2}.

\noindent \textbf{Ranking}: Similar to the ASR task, the final ranking is determined by averaging the ranks across two primary metrics: UTMOS and bitrate. UTMOS is ranked in descending order, while bitrate is ranked in ascending order. To allow different focuses on sampling rates, we separate the bitrate into two groups ($16$kHz and $48$kHz), depending on the sampling rate of the resynthesized waveform. The ranking of both UTMOS and bitrate would be considered separately in each group. If there's a tie in the overall ranking, UTMOS rankings will serve as a tiebreaker to establish the final positions.

\subsection{TTS (Acoustic + Vocoder) task}
\label{sec: tts}

In the challenge, the TTS (Acoustic + Vocoder) track focuses on the use of discrete units as an intermediate representation in a cascaded TTS system. Here the cascaded TTS highlights the TTS system that consists of both an acoustic model and a vocoder. This approach is supported by several research findings suggesting that discrete representations offer considerable benefits for speech synthesis systems. The potential benefits include easy predictability, stability during training, and versatility in interacting with different modalities~\cite{yan-etal-2023-espnet, barrault2023seamless, maiti2023voxtlm, wang2023neural, yang2023uniaudio}. Participants are encouraged to explore the use of discrete units to enhance both the performance and efficiency of TTS systems.

\noindent \textbf{Task Definition}: The challenge's TTS task involves converting text into speech signals. Participants are required to use a cascaded TTS system where the acoustic model translates text into discrete units $\mathbf{U}$, and the vocoder converts $\mathbf{U}$ into the predicted waveform $\hat{\mathbf{x}}$. The model type or size for both the acoustic model and the vocoder is not subject to any limitations.

\noindent \textbf{Data}: The challenge focuses on a single-speaker TTS task using the LJSpeech dataset~\cite{ljspeech17}, with $250$ utterances set aside for both development and test purposes. While there are no restrictions on the data used for learning or extracting discrete units, the provided training data must exclusively be used for training the TTS system components.

\noindent \textbf{Evaluation metrics}: The evaluation includes the same four metrics as in the TTS (Vocoder) track, with the addition of the word error rate (WER) from Whisper-large V2~\cite{radford2023robust}.

\noindent \textbf{Ranking}: The ranking methodology mirrors that of the TTS (Vocoder) track, focusing on a combined assessment of speech quality and discrete unit efficiency to determine the overall performance standings. In case of a tie in the overall ranking, UTMOS ranking will be the tiebreaker for final positions.

\subsection{SVS track}
\label{ssec:svs_track}

The SVS track distinguishes itself from TTS by focusing on the intersection of music and speech processing. Unlike previous works that often extend TTS frameworks to SVS~\cite{lu2020xiaoicesing, VISinger, yang2023uniaudio}, this challenge treats singing synthesis as an independent track to foster deeper exploration into singing-specific features.

\noindent \textbf{Task Definition}: Singing synthesis entails generating singing voices using musical score information. Mirroring the TTS (Acoustic + Vocoder) track, this challenge adopts a cascaded approach, incorporating an acoustic model and a vocoder. The acoustic model's role is to translate the music score into a sequence of discrete units $\mathbf{U}$, while the vocoder is tasked with synthesizing the waveform from $\mathbf{U}$. There are no additional constraints imposed on SVS modeling for this challenge.

\noindent \textbf{Data}: For the SVS track, the dataset employed is the $5.2$-hour single-singer Opencpop dataset~\cite{wang22b_interspeech}. The challenge adheres to the original dataset's train, development, and test splits. Similar to the TTS tracks, training for the SVS track must only utilize the provided dataset, although any data source is permissible for extracting discrete representations.

\noindent \textbf{Evaluation Metrics}: The evaluation for the SVS track encompasses four metrics: MCD, F0 RMSE, MOS, and bitrate. The objective metrics (MCD, F0 RMSE, and bitrate) follow the same calculation methodology as in the TTS tracks. For MOS, $20$ subjects rate the submissions (i.e., $206$ utterances) on a $5$-point scale, with $1$ indicating "unreasonable singing" and $5$ denoting "natural singing comparable to human performance."

\noindent \textbf{Ranking}: The overall ranking is determined by averaging the ranks across two key metrics: MOS and bitrate. MOS rankings are in descending order, while bitrate rankings are in ascending order. In the event of tied rankings, priority is given to the MOS results for final ranking decisions.

\section{Baseline Systems}
\label{sec:baseline}

\subsection{ASR baseline}
\label{ssec:asr_exp}
The baseline system follows the model used in~\cite{chang2023exploration} and is implemented using ESPnet~\cite{watanabe2018espnet}. The ASR backbone uses the joint CTC/attention-based encoder-decoder architecture based on the E-Branchformer~\cite{kim2023branchformer}\footnote{We follow the model configurations in \scriptsize{\url{https://github.com/espnet/espnet/blob/master/egs2/interspeech2024_dsu_challenge/asr2/conf/tuning/train_discrete_asr_e_branchformer1_1gpu_lr5e-4_warmup30k.yaml}}}.
The baseline model undergoes training for 100 epochs, utilizing a single Nvidia V-100 32GB GPU, with a total training time of about 18 hours.

For discrete speech units, $1,024$-dimensional features are extracted from the $21$-st layer of the WavLM-Large~\cite{chen2022wavlm} model. A $k$-means model with $2,000$ clusters is trained using randomly chosen $15\%$ of the data from the training set, described in Section.~\ref{ssec:asr_track}. Additionally, repeated tokens are removed, and the BPE model is applied with a vocabulary size of $6,500$, i.e. $|\mathcal{V}|=6,500$ in Section.~\ref{ssec:asr_track}.

\subsection{TTS baseline}
\label{ssec:tts_exp}
\noindent \textbf{Vocoder track}: For the TTS (Vocoder) track, the baseline involves $k$-means ($k=|\mathcal{V}|=500$) clustering over the whole training set on the 9\textsuperscript{th} layer outputs of a pre-trained HuBERT-base model~\cite{hsu2021hubert}. The setting is aligning with previous SSL-based unit extraction methodologies~\cite{polyak21_interspeech, lee2022direct, shi2023enhancing, maiti2023voxtlm, barrault2023seamless, nguyen2023expresso, yan-etal-2023-espnet}. The derived token sequence is processed using a discrete-token-based HiFi-GAN within the ESPnet framework~\cite{yan-etal-2023-espnet, hayashi2020espnet, hayashi2021espnet2}.\footnote{We follow the model configurations in \scriptsize{\url{https://github.com/kan-bayashi/ParallelWaveGAN/blob/master/egs/cvss_c/hubert_voc1/conf/hifigan_hubert_duration.v1.yaml}}}

\noindent \textbf{Acoustic + Vocoder track}: For the TTS (Acoustic + Vocoder) track, we separately train an acoustic model and a vocoder. For the vocoder, we use the same vocoder setting as the TTS (Vocoder) track with LJSpeech training data. For the acoustic model, we adopt a Fastspeech2 architecture~\cite{ren2020fastspeech}, adapted to output discrete units instead of spectrograms. The acoustic model configuration follows the LJSpeech Fastspeech2 recipe in ESPnet-TTS~\cite{hayashi2020espnet}.\footnote{\scriptsize{\url{https://github.com/espnet/espnet/blob/master/egs2/ljspeech/tts1/conf/tuning/train_fastspeech2.yaml}}}

\subsection{SVS baseline}
\label{ssec:svs_exp}

The SVS baseline consists of an acoustic model and a vocoder. The acoustic model is adapted from XiaoiceSing~\cite{lu2020xiaoicesing}. We replace the output spectrogram in original XiaoiceSing into two streams, including one stream of quantized fundamental frequency (with a resolution of 10Hz) and another stream with semantic discrete tokens. The discrete tokens are extracted from the 6\textsuperscript{th} layer of WavLM-large with a $k$-means ($k=|\mathcal{V}|=1024$) over the whole training set. The acoustic model consists of an encoder, a length regulator and a decoder. The implementation is based on ESPnet-Muskits~\cite{shi2022muskits}. The network architecture and the training configuration follow the XiaoiceSing model configuration in corresponding Opencpop recipe.\footnote{\scriptsize{\url{https://github.com/espnet/espnet/blob/master/egs2/opencpop/svs1/conf/tuning/train_xiaoice.yaml}}} The vocoder utilizes the same architecture as the TTS baselines.

\section{Preliminary Results}
\label{sec:results}
This section presents the initial results that we collected before the paper deadline. Due to time constraints, a more in-depth analysis and detailed results will be presented following the conclusion of the challenge.

\subsection{ASR results}
\label{ssec: asr_results}

Prior to the submission deadline, nine systems were submitted for the ASR track. We list the performance of the top-$3$ submitted systems in Table~\ref{tab:asr_table}. In comparison to the provided baseline system (B1), the submitted system S2 outperformed on all metrics. S2 employed a similar discrete token process as the baseline, utilizing the XLSR2-300M for feature extraction. A 2000-cluster $k$-means model was applied to the features, followed by BPE with a vocabulary size of 6000. This approach resulted in a notable $7\%$, $23\%$, and $26\%$ reduction in $\text{CER}_\text{EN}$, $\text{CER}_\text{ML}$, and bitrate, respectively. S1 and S3 utilize the fusion techniques to combine the discrete tokens from multiple streams.  In contrast to other approaches, S3 employs the Hidden Markov Model (HMM) for computing the discrete tokens.

\begin{table}[!tbp]
    \centering
    \caption{The performance of the baseline and submitted systems on the ASR task. We use CERs on the English test sets and the multi-lingual counterpart, as well as the bitrate. Brief discrete token information collected from the participants is added.}
    \vspace{-5pt}
    \resizebox{\linewidth}{!}{
    \begin{tabular}{c|c|c|c|c}
    \toprule
    Team ID & Discrete token info & $\text{CER}_\text{EN}$ & $\text{CER}_\text{ML}$ & Bitrate \\
    \midrule
    B1     & WavLM\_Large\_21st, $k$-means & 2.37 & 22.40 & 356.19 \\
    \midrule
    S1     & (XLSR2\_300M, WavLM\_Large), $k$-means & 1.91 & 16.03 & 946.77 \\
    S2     & XLSR2\_300M, $k$-means & 2.21 & 17.32 & 262.64 \\
    S3     & (WavLM\_Large, XLS-R), HMM-GMM & 1.98 & 20.23 & 599.20 \\
    \bottomrule
    \end{tabular}
    \label{tab:asr_table}
    }
    \vspace{-5pt}
\end{table}

\subsection{TTS results}
\label{ssec: tts_results}

For this track, we received 13 systems for the TTS~(Vocoder) task and 8 systems for the TTS (Acoustic + Vocoder) task. In this paper, we present the preliminary results by selecting the top three systems in terms of the overall ranking. 

For the TTS (Vocoder) track, all six top systems from both 16kHz and 48kHz settings surpass the baseline by a large margin in UTMOS score and other objective evaluation metrics, suggesting better resynthesis quality. 
Model S1 refines the SSL pre-trained representation with audio resynthesis tasks, and shows the best UTMOS score in the challenge. 
Different from B1 and S1 originated from SSL pre-trained models, the other two methods S2 and S3 adapt neural codec-based models, including Descript Audio Codec (DAC)~\cite{kumar2024high} and APCodec~\cite{ai2024apcodec}. The discrete unit extractor is optimized on the audio resynthesis task with adversarial training. The codebooks from the pre-trained codecs are then used as the discrete representation for the task. Notably, their bitrates are generally higher than B1 and S1 due to the use of multi-stream information.

In the TTS (Acoustic + Vocoder) track, we compare the top three models: S1 and S2 employ discrete representations from a DAC-based neural codec, while S3 utilizes explicit vector quantization within an end-to-end TTS training framework. S3 stands out by delivering the highest UTMOS scores and the lowest WER, illustrating its superior performance. However, this comes at the expense of a higher bitrate. Conversely, S1 achieves a commendable equilibrium between bitrate efficiency and UTMOS performance, presenting a viable option for scenarios where a balance between audio quality and resource usage is essential.

\begin{table}[!tbp]
    \centering
    \caption{The performance of the baseline and submitted systems on the TTS (Vocoder) task. $S$ is the sampling rate of targeted audio from the system.}
    \vspace{-5pt}
    \resizebox{0.95\linewidth}{!}{
    \begin{tabular}{c|c|c|c|c|c}
    \toprule
    Team ID & $S$ &  MCD & F0 RMSE & UTMOS & Bitrate \\
    \midrule
    B1     & 16k & 7.19 & 0.42 & 2.27 & \textbf{448.3}  \\
    \midrule
    S1     & 16k & 6.24 & 0.24 & \textbf{3.59} & 547.0 \\
    S2     & 24k & 4.81 & 0.21 & 3.58 & 670.3 \\
    S3     & 16k & 3.57 & \textbf{0.18} & 3.57 & 1479.5 \\
    \midrule
    S4     & 48k & \textbf{3.54} & \textbf{0.18} & 3.56 & 1479.5 \\
    S5     & 48k & 4.47 & \textbf{0.18} & 3.48 & 834.0 \\
    S6     & 48k & 4.47 & \textbf{0.18} & 3.48 & 834.0 \\
    \bottomrule
    \end{tabular}
    \label{tab:asr_table}
    }
    \vspace{-5pt}
\end{table}

\subsection{SVS results}
\label{ssec: tts_results}

For the SVS challenge, six systems were submitted. We focus on the top-$3$ systems based on their performance metrics. S1 and S2 employ SSL-based discrete tokens within a non-autoregressive framework, contrasting with S3, which is built on neural codecs and operates in an autoregressive manner. Despite the varied configurations among the models, S1 and S2, along with baseline B1, outperform S3. This superiority could be attributed to the limited training data provided in the challenge, introducing  additional challenges to autoregressive modeling. Though the data scarciy is a common constraint in SVS tasks, the SSL-based discrete units utilized in S1 and S2 appear to offer robust representations for discrete SVS systems.

\begin{table}[!tbp]
    \centering
    \caption{The performance of the baseline and submitted systems on the TTS (Acoustic + Vocoder) task.}
    \vspace{-5pt}
    \resizebox{0.95\linewidth}{!}{
    \begin{tabular}{c|c|c|c|c|c}
    \toprule
    Team ID & MCD & F0 RMSE & WER & UTMOS & Bitrate \\
    \midrule
    B1     & 7.19 & \textbf{0.26} & 8.1 & 3.73 & 448.3 \\
    \midrule
    S1     & \textbf{6.96} & 0.29 & 7.7 & 4.33 & \textbf{277.6}  \\
    S2     & 7.15 & 0.29 & 7.4 & 4.33 & 353.9  \\
    S3     & 7.70 & 0.29 & \textbf{6.8} & \textbf{4.42} & 727.5 \\
    \bottomrule
    \end{tabular}
    \label{tab:tts_table}
    }
    \vspace{-5pt}
\end{table}

\begin{table}[!tbp]
    \centering
    \caption{The performance of the baseline and submitted systems on the SVS task. 95\% confidence interval is in parentheses.}
    \vspace{-5pt}
    \resizebox{0.9\linewidth}{!}{
    \begin{tabular}{c|c|c|c|c}
    \toprule
    Team ID & MCD & F0 RMSE & MOS & Bitrate \\
    \midrule
    B1     & 8.47 & 0.18 & 3.43 ($\pm$ 0.05) & 2094.7 \\
    \midrule
    S1     & \textbf{7.56} & \textbf{0.17} & \textbf{3.70} ($\pm$ 0.05) & 1899.9 \\
    S2     & 7.72 & 0.19 & 3.09 ($\pm$ 0.06) & 874.8 \\
    S3     & 11.44 & 0.24 & 2.73 ($\pm$ 0.06) & 725.9 \\
    \bottomrule
    \end{tabular}
    \label{tab:svs_table}
    }
    \vspace{-5pt}
\end{table}

\section{Conclusion}
\label{sec:conclusion}
This paper serves as a comprehensive overview of the Interspeech 2024 challenge on speech processing with discrete units. The challenge garnered a notable 40 submissions across ASR, TTS, TTS-vocoder, and SVS tasks, underscoring the significant interest in this domain. We provide detailed insights into the motivation, challenge rules, baseline systems, and initial submission results.

Upon reviewing the initial submissions, several initial observations can be made. Notably, in the ASR task, the utilization of semantic tokens from SSL models demonstrates promising outcomes. While for TTS tasks, neural codec-based model usually exhibit high-quality acoustics, which significantly enhance the synthesized audio quality. In the SVS track, on the other hand, SSL-based units demonstrate strong performance over the dataset, suggesting the rich acoustic information can be also obtained from SSL-based pre-trained models in the singing domain. However, to derive more nuanced and conclusive findings, a thorough and in-depth analysis requires additional time and efforts. Detailed analyses and findings will be unveiled as we invest the necessary resources in their examination.

\section{Acknowledgements}
Experiments of this work used the Bridges2 system at PSC and Delta system at NCSA through allocations CIS210014 and IRI120008P from the Advanced Cyberinfrastructure Coordination Ecosystem: Services \& Support (ACCESS) program, supported by NSF grants \#2138259,\#tel:2138286, \#tel:2138307, \#tel:2137603, and \#tel:2138296.

\bibliographystyle{IEEEtran}
\bibliography{mybib}

\end{document}